\documentclass[sigconf,screen]{acmart}

\usepackage{adjustbox}
\usepackage{mdframed}
\usepackage{tabularx}
\usepackage{multirow}
\usepackage{listings}
\usepackage{tcolorbox}
\usepackage{ragged2e} 
\usepackage{subfig}
\usepackage[dvipsnames]{xcolor}
\usepackage{microtype} 

\newcommand{\rev}[1]{\textcolor{Black}{#1}}
\newcolumntype{Y}{>{\Centering\arraybackslash}X}

\setcopyright{cc}
\setcctype{by}
\acmDOI{10.1145/3832783.3834523}
\acmYear{2026}
\copyrightyear{2026}
\acmISBN{979-8-4007-2882-2/2026/10}
\acmConference[ASE '26]{Proceedings of the 41st IEEE/ACM International Conference on Automated Software Engineering}{October 12--16, 2026}{Munich, Germany}
\acmBooktitle{Proceedings of the 41st IEEE/ACM International Conference on Automated Software Engineering (ASE '26), October 12--16, 2026, Munich, Germany}
\acmSubmissionID{ase26ind-p302-p}
\received{2026-05-01}
\received[accepted]{2026-07-01}

\begin{document}








\title{AgenticSCR: An Autonomous Agentic Secure Code Review for Immature Vulnerabilities Detection}

\author{Wachiraphan Charoenwet}
\orcid{0000-0002-9814-3514}
\affiliation{%
  \institution{University of Melbourne}
  \city{Parkville}
  \country{Australia}
}
\email{wcharoenwet@student.unimelb.edu.au}

\author{Kla Tantithamthavorn}
\correspondingauthor
\orcid{0000-0002-5516-9984}
\affiliation{%
  \institution{Monash University}
  \city{Clayton}
  \country{Australia}
}
\email{chakkrit@monash.edu}

\author{Patanamon Thongtanunam}
\orcid{0000-0001-6328-8839}
\affiliation{%
  \institution{University of Melbourne}
  \city{Parkville}
  \country{Australia}
}
\email{patanamon.t@unimelb.edu.au}

\author{Hong Yi Lin}
\orcid{0009-0004-5368-8897}
\affiliation{%
  \institution{University of Melbourne}
  \city{Parkville}
  \country{Australia}
}
\email{holin2@student.unimelb.edu.au}

\author{Minwoo Jeong}
\orcid{0009-0008-9960-1741}
\affiliation{%
  \institution{Atlassian}
  \city{Bellevue}
  \country{USA}
}
\email{mjeong@atlassian.com}

\author{Ming Wu}
\orcid{0000-0002-6993-1431}
\affiliation{%
  \institution{Atlassian}
  \city{Bellevue}
  \country{USA}
}
\email{mwu2@atlassian.com}

\begin{abstract}
  
  

  Secure code review is critical during pre-integration, where Atlassian developers rely on lightweight analysis tools, while deep security assessment is deferred to later stages, delaying feedback and increasing remediation costs.
  Existing static analyzers are often noisy and struggle with context-dependent or partially manifested vulnerabilities, while static large language model (LLM) reviewers are constrained by context windows and lack tool interaction.
  Agentic AI, which combines LLMs with code navigation, shows promise; however, its effectiveness for early-stage secure code review remains underexplored.
  We present AgenticSCR, an agentic secure code reviewer augmented with security-focused semantic memory that grounds reasoning in structured security knowledge to detect vulnerabilities before they fully manifest.
  AgenticSCR achieves at least 153\% relative improvement in generating comments with correct localization, vulnerability type, and relevance over static LLM baseline, multi-agent reviewer, and SAST tools. 
  In a shadow deployment, 54\% of its comments were validated by security engineers for developer reporting, demonstrating practical utility while underscoring the difficulty of the task.
  These findings position the security-focused semantic memory as a promising direction for agentic secure code review, enabling early-stage vulnerability identification.
  Our approach builds an important step toward reliable localization, detection, and explanation in shift-left security practices.
  

\end{abstract}



\maketitle

\begin{figure}[h]
    \centering
    \includegraphics[width=\linewidth]{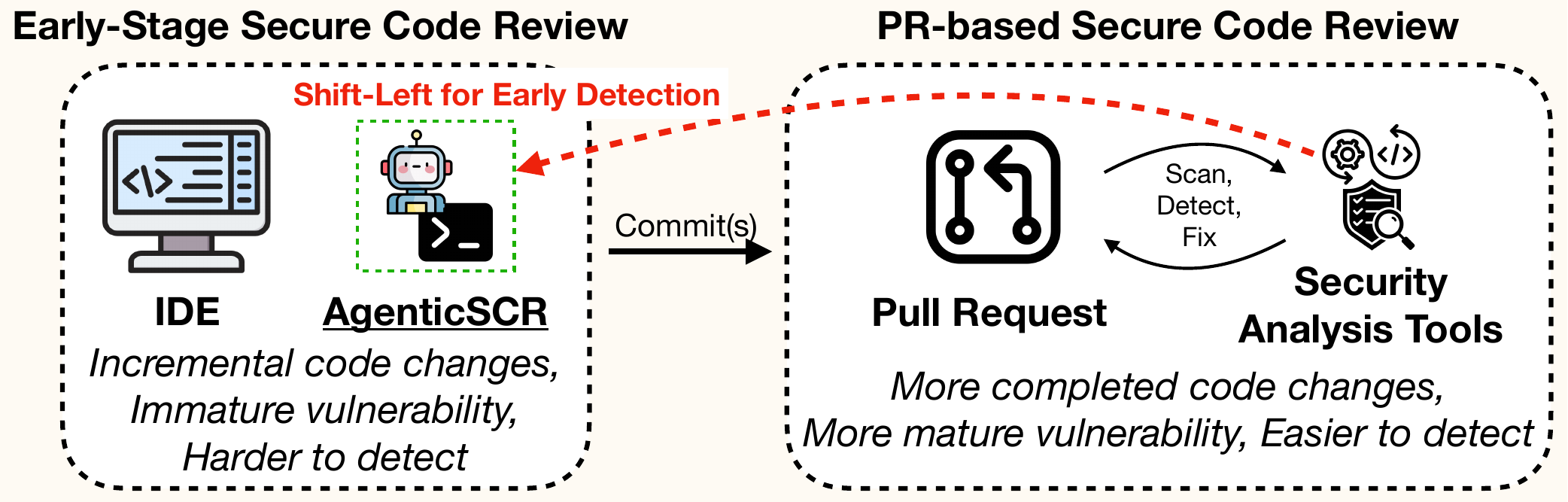}
    \caption{A visual comparison between early-stage secure code review (before code is pushed) and pull request–based secure code review practices.}
    \label{fig:precommit_vs_pullrequest}
\end{figure}

\section{Introduction}

Secure code review is a fundamental practice for preventing vulnerabilities from reaching production systems.
Among the various review stages (Figure~\ref{fig:precommit_vs_pullrequest}), \emph{early-stage secure code review}—where developers inspect their own changes before pushing to a shared repository—offers an opportunity for low-cost intervention, aligning with the principles of shift-left security~\cite{Baum2016ComparingSimulation}.
In practice, Atlassian developers often rely on automated tools such as static analyzers at this stage, while deeper security assessment by experts is deferred to later phases (e.g., post-integration), delaying feedback and increasing remediation costs.
A key challenge is that many issues introduced at this stage are \emph{\textbf{immature vulnerabilities}}: incomplete, latent, or context-dependent weaknesses introduced through small incremental changes~\cite{Nguyen2025,Tahaei2021SecurityThem}.
These changes may appear benign in isolation, yet evolve into exploitable vulnerabilities as surrounding code is added~\cite{Nguyen2023,lomio2022jitvd}.
Prior work shows that developers frequently overlook such security-relevant issues during review, particularly when changes are small, and their broader implications are difficult to reason about~\cite{Braz2021Why-,Braz2022SoftwarePerspective,Paul2021WhyProject,Roy2024ExploringAnalysis,fu2023vulexplainer}.


Existing techniques struggle under these conditions. 
Static analyzers dominate early-stage security checks due to their low latency and rule-based design, but they are primarily tuned for mature vulnerabilities in the complete code~\cite{Charoenwet2024AnReview,panichella2015saner}.
Consequently, when applied to early-stage changes, they often produce noisy alerts while missing context-dependent weaknesses, leading developers to ignore them~\cite{Charoenwet2024AnReview,panichella2015saner}.
Large language models (LLMs) have shown promise for code review~\cite{Liu2025SecureReviewer:Fine-tuning,Peng2025ICodeReviewer:Prompts,HgtJIT2023,AIBugHunter2023,fu2023chatgpt,fu2022vulrepair,yu2025preliminarystudylargelanguage}, but static LLM reviewers lack repository-level context and tool interaction, limiting their ability to reason about immature vulnerabilities.

\rev{In this work, we present \textsc{\textbf{AgenticSCR}}, an agentic secure code review support framework that surfaces candidate security issues during early-stage code review.}
\rev{Unlike prior work applying static LLM prompting without repository context or multi-agent review without security objectives, AgenticSCR combines LLM reasoning with repository navigation, tool invocation, and security-focused semantic memory~\cite{sumers2023cognitive}.}
It is structured as detector and validator subagents: the \emph{detector} localizes and explains potential vulnerabilities, while the \emph{validator} filters noisy findings.
Specifically, the detector is augmented with semantic memory derived from Static Application Security Testing (SAST) rules, and the validator employs a Common Weakness Enumeration (CWE)-based validation framework.

To evaluate AgenticSCR, we curate \textsc{\textbf{SCRBench}}, a repository-aware, human-annotated, line-level benchmark of immature vulnerabilities in early-stage code changes.
AgenticSCR correctly localizes, identifies, and explains vulnerabilities in 17.5\% of generated comments, outperforming a static LLM-based baseline~\cite{Liu2025SecureReviewer:Fine-tuning} by 10.6 percentage points and a multi-agent reviewer~\cite{tang-etal-2024-codeagent} by 12.3 percentage points, while producing only 32\% and 21\% as many comments, respectively.
Ablation analysis shows that augmenting both detector and validator with security knowledge yields the best performance, improving accuracy by 10.2 percentage points over a vanilla agentic baseline while reducing comment volume by 81\%.
In a shadow deployment, 54\% of AgenticSCR’s comments were validated by Atlassian product security engineers and reported to developers; 27.3\% of reported comments have already \rev{been addressed by developers}, with the remainder under consideration.
Collectively, these findings demonstrate the utility of security-augmented semantic memory for agentic secure code review and highlight its potential to improve early-stage vulnerability detection in industrial workflows.

\noindent\textbf{Contributions.} This paper makes the following contributions:
\begin{itemize}
\setlength{\leftskip}{-10pt} 
    \item \rev{We present AgenticSCR, an agentic secure code review support framework that augments repository-aware reasoning with security-focused semantic memory to surface immature vulnerabilities.}
    \item \rev{We construct \textsc{SCRBench}, a repository-aware, human-annotated, line-level benchmark of immature vulnerabilities in early-stage code changes, addressing the lack of repository-level context and line-level annotations in existing benchmarks.}
    \item We evaluate AgenticSCR with secure review tasks across three metrics—line-level localization, vulnerability type identification, and secure review comment generation—and demonstrate substantial improvements over static LLM and multi-agent baselines, and SAST tools.

\end{itemize}


\section{Background and Related Work}
\label{section:background}


\textbf{Secure code review} is a cornerstone of modern secure software development lifecycles. 
It is embedded across multiple stages of development, supported by a growing ecosystem of automated tools. 
At Atlassian,\footnote{\url{https://www.atlassian.com/blog/add-ons/code-review-best-practices}} industrial practices typically follow a multi-stage pipeline that progressively increases review scope and cost (see Figure~\ref{fig:precommit_vs_pullrequest}), e.g., \emph{early-stage secure code review $\Rightarrow$ Pull Request (PR)-based secure code review}.
In \emph{early-stage secure code review}, developers typically review their own code changes to identify potential issues before integration.

A key challenge at this stage is that vulnerabilities may be highly context-dependent~\cite{Bojanova2016TheBugs} or latent across multiple editing steps.
For example, insecure API usage, missing validation logic, or flawed authorization checks may appear benign in isolation but evolve into exploitable weaknesses as surrounding code is added~\cite{lomio2022jitvd, Nguyen2023}.
These \textbf{\emph{immature vulnerabilities}} are therefore difficult to detect using techniques designed for post-hoc analysis of finalized code, where issues are already fully manifested~\cite{Nguyen2025, HgtJIT2023, JITLine2021}.
As a result, many remain undetected and propagate into main repositories.

This phenomenon is observed in real-world development.
For example, a code change\footnote{\url{https://github.com/saltstack/salt/commit/aa87d67258a5b6742fc53d06c7bdac0f643bc9f1}}
in SaltStack, an automation engine, introduces validation logic intended to prevent directory traversal but uses an incorrect regular expression.
A subsequent change\footnote{\url{https://github.com/saltstack/salt/commit/0976f8f7131975a1ae29b2724069a301a870a46d}}
attempts to strengthen the validation, yet fails to enforce it consistently, partially mitigating the issue while leaving the vulnerability latent and difficult to identify from either individual change.
This example illustrates how incremental changes that appear benign or corrective can collectively form a real vulnerability~\cite{CVE-2017-14695}.

To support early-stage secure code review, developers often rely on warnings from Static Application Security Testing (SAST) tools integrated into modern Integrated Development Environments (IDEs) or Command-Line Interfaces (CLIs).
These tools are designed for low latency and frequent execution using predefined rules and vulnerability patterns.
However, prior studies~\cite{Charoenwet2024AnReview, panichella2015saner} and industry practitioners consistently report that SAST tools generate substantial noise, including false positives and low-severity warnings.
As a result, developers frequently ignore or disable these checks, undermining their preventive value.


Recent work has begun to explore richer Deep Learning (DL)-driven analyses within IDEs to bridge this gap. 
Prior work performs just-in-time vulnerability detection at the line~\cite{JITLine2021, fu2022linevul, Peng2025ICodeReviewer:Prompts, Sun2025BitsAI-CR:Practice}, commit~\cite{CodeFuseCRBench2025, Liu2025SecureReviewer:Fine-tuning, Cihan2025EvaluatingReview, Haider2024PromptingGeneration}, file~\cite{AIBugHunter2023}, and pull-request levels~\cite{Baciejowski2023}, typically using monolithic LLM predictors or classifiers.
However, these methods often operate on narrow views of code (e.g., diffs or small code blocks) without considering broader repository context.
Additionally, some approaches depend on large proprietary fine-tuning datasets and restrictive licenses, limiting enterprise and on-premise adoption.
Consequently, commit- and PR-level analyses remain largely reactive, detecting issues after integration rather than preventing them during early-stage development.

AIBugHunter~\cite{AIBugHunter2023} is a closely related tool for shift-left secure checks, integrating deep learning-based vulnerability detection into Visual Studio Code with real-time, line-level localization, classification, and repair suggestions. 
However, it analyzes entire open files and is optimized for continuous file-level scanning rather than incremental changes across the repository. 
Its reliance on frequent model invocations can also introduce inefficiency and noise, limiting its effectiveness for early-stage secure code review.

\textbf{Related Work and Limitations.}
Despite rapid progress in LLM-based code review~\cite{Liu2025SecureReviewer:Fine-tuning, Cihan2025EvaluatingReview, Haider2024PromptingGeneration, Peng2025ICodeReviewer:Prompts, Sun2025BitsAI-CR:Practice, tang-etal-2024-codeagent, ZhangLaura2025}, existing approaches still fall short of the requirements of early-stage secure code review.
Table~\ref{table-related-work} compares AgenticSCR with existing LLM-based code review approaches.
Particularly, we identify the following limitations:

\begin{table}[t]
\centering

\scriptsize   
\caption{The key differences between this and prior work.}

\begin{tabularx}{\linewidth}{
            >{\hsize=1.5\hsize}X
            >{\hsize=0.6\hsize}Y
            >{\hsize=0.5\hsize}Y
            >{\hsize=1.3\hsize}Y
            >{\hsize=0.8\hsize}Y
            >{\hsize=1.2\hsize}Y
          }
\toprule
\textbf{Related Work} & 
\textbf{Approach} & 
\textbf{Security-Focus} & 
\textbf{Dataset} & 
\textbf{Evaluation Granularity} & 
\textbf{Additional Context/ Knowledge}\\
\toprule
\textbf{\citet{Liu2025SecureReviewer:Fine-tuning}} & LLM & \textbf{Yes} & CodeReviewer & Diff & 
- \\   
\textbf{\citet{Cihan2025EvaluatingReview}} &  LLM & No & HumanEval & Block & 
-  \\   
\textbf{\citet{Haider2024PromptingGeneration}} &  LLM & No & CodeReviewer & Diff & 
- \\   
\textbf{\citet{Peng2025ICodeReviewer:Prompts}} &  LLM & \textbf{Yes} & Private & \textbf{Line} &
CWE \\   
\textbf{\citet{Sun2025BitsAI-CR:Practice}} &  LLM & Partially & Private & \textbf{Line} &
-  \\   

\textbf{\citet{ZhangLaura2025}} &  LLM & No & LAURA & \textbf{Line} &
AST  \\   

\textbf{\citet{tang-etal-2024-codeagent}} & \textbf{Multi-Agents} & No & TransReview, AutoTransform, T5-Review & \textbf{Diff} & 
-  \\   


\midrule
\textbf{AgenticSCR} & \textbf{Agentic AI} & \textbf{Yes} & \textbf{SCRBench} & \textbf{Line} &
\textbf{Repository, SAST Rules, \& CWE} \\   
 
\bottomrule
\end{tabularx}
\label{table-related-work}
\vspace{-5mm}
\end{table}

\begin{itemize}
\setlength{\leftskip}{-10pt} 
\item \textbf{Lack of contextual information within the repository:}
Most existing approaches~\cite{Liu2025SecureReviewer:Fine-tuning, Cihan2025EvaluatingReview, Haider2024PromptingGeneration, Peng2025ICodeReviewer:Prompts, Sun2025BitsAI-CR:Practice, tang-etal-2024-codeagent, ZhangLaura2025} rely on \emph{static LLM prompting}—a pre-defined prompt applied uniformly across inputs—using only PR-level information (e.g., code diffs, issue descriptions).
This limits their ability to adaptively explore and incorporate broader repository context.
Because early-stage vulnerabilities are highly context-dependent~\cite{Bojanova2016TheBugs}, identifying such weaknesses requires reasoning over repository-level context and recognizing patterns that may not yet be fully manifested.

\item \textbf{Lack of Ground-Truths and Benchmark for Line-Level Secure Code Review.}
Existing vulnerability datasets~\cite{Liu2025SecureReviewer:Fine-tuning, TufanoCR2021, BhandariCVEFixes2021, FanBigVul2020, CodeBERT2020} focus on file- or function-level detection of fully manifested vulnerabilities, primarily for prediction or classification.
Similarly, existing code review benchmarks~\cite{Liu2025SecureReviewer:Fine-tuning, Cihan2025EvaluatingReview, Haider2024PromptingGeneration, tang-etal-2024-codeagent} operate at the diff-hunk level.
They lack high-quality line-level annotations and do not capture \emph{immature vulnerabilities} in early-stage code, making them unsuitable for evaluating pre-commit secure code review.
Although recent work~\cite{Peng2025ICodeReviewer:Prompts, Sun2025BitsAI-CR:Practice} explores line-level review, their datasets are not publicly available.
Moreover, most datasets omit repository-level context (e.g., dependencies and configuration files), which is critical for detecting early-stage vulnerabilities.

\end{itemize}

\rev{These gaps motivate us to empirically investigate how security-focused semantic memory augmentation shapes agentic reasoning for early-stage secure code review—a setting that existing benchmarks and approaches do not address.}

\section{AgenticSCR: an Autonomous Agentic Secure Code Reviewer}
\begin{figure*}[t]
    \centering
    \includegraphics[width=\linewidth]{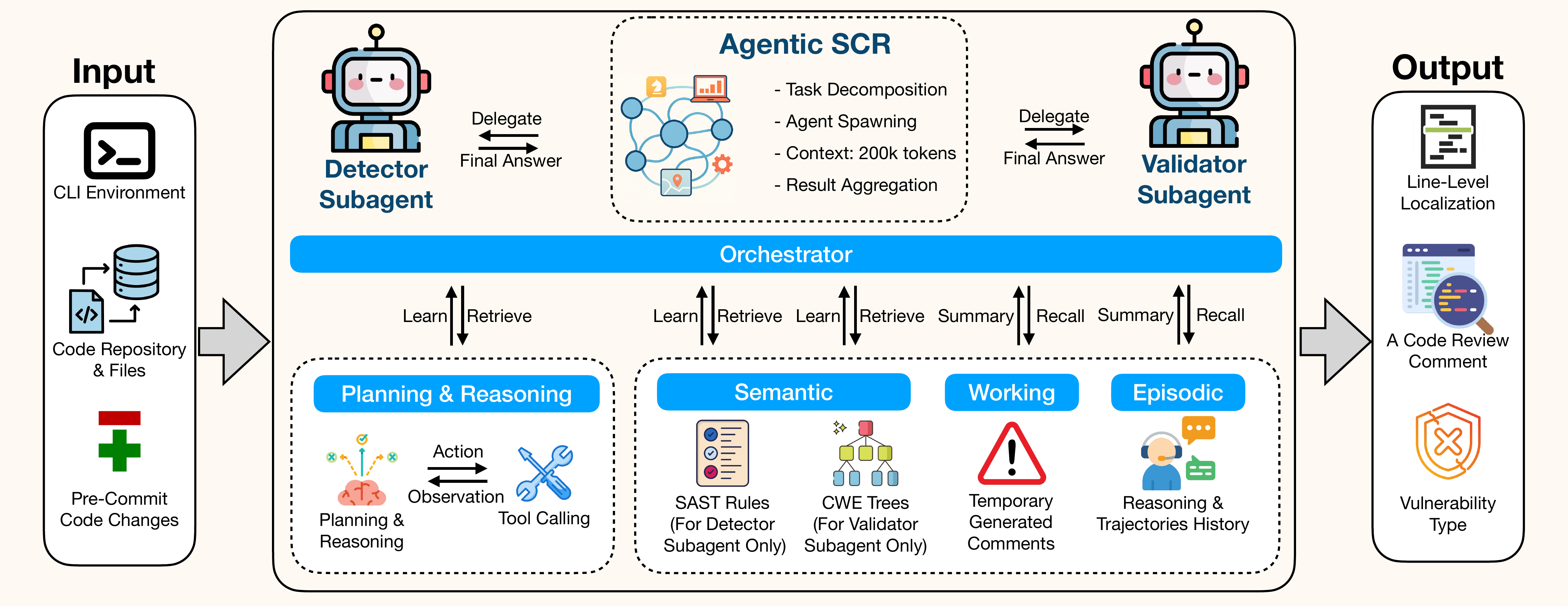}
    \caption{An overview of the architecture of the Agentic Secure Code Review workflow.}
    \label{fig:agenticscr}
\end{figure*}

In this section, we present the architecture of our Autonomous Agentic Secure Code Review framework (\textsc{AgenticSCR}).
Figure~\ref{fig:agenticscr} presents an overview of the Agentic Secure Code Review workflow. 

\subsection{Overview}

\rev{AgenticSCR is an agentic secure code review support framework, designed to operate in command-line environments, where developers iteratively make and refine code changes, and assist developers by surfacing candidate security issues before integration.}
At this stage, code changes are often incomplete and involve subtle, context-dependent vulnerability patterns that require iterative reasoning and targeted exploration. 
To address this challenge, AgenticSCR adopts an agentic AI paradigm~\cite{sapkota2025agentic}, overcoming limitations of static analyzers and monolithic LLM-based approaches, which often lack contextual awareness, adaptability, and persistence across iterations. 
AgenticSCR decomposes high-level security goals into specialized subtasks, coordinates subagents, explores repository-level context beyond code diffs, and adapts its strategy based on intermediate findings. 
Its memory mechanisms enable continuity across analysis steps, reducing redundant effort and supporting progressively refined decision-making.
We formally defined AgenticSCR as follows.
\[
\text{AgenticSCR} = \langle \mathcal{E}, \mathcal{A}, \mathcal{G}, \mathcal{M}, \mathcal{T},  \Pi \rangle
\]
where:
\begin{itemize}
\setlength{\leftskip}{-10pt} 
    \item $\mathcal{E}$ denotes the environment, including a CLI, the repository file system with early-stage code changes, and toolchains.
    \item $\mathcal{A} = \{a_d, a_v\}$ is a set of coordinated subagents: the \emph{detector} $a_d$ for localizing, detecting, and explaining vulnerabilities, and the \emph{validator} $a_v$ for validating them.
    \item $\mathcal{G}$ denotes high-level goals, including localization, detection, and explanation of vulnerabilities.
    \item $\mathcal{M}$ is a structured memory system that supports adaptive reasoning and continuity across iterations, inspired by cognitive architectures~\cite{sumers2023cognitive}.
    \item $\mathcal{T}$ is a set of executable tools (e.g., code expansion, diff inspection, bash) enabling interaction with the environment.
    \item $\Pi$ is a coordination policy governing task decomposition, agent collaboration, and tool selection.
\end{itemize}


We describe each of the components below.

\subsection{Dynamic Context Building Environments for Input Perception ($\mathcal{E}$)}


The environment ($\mathcal{E}$) defines the external context in which \textsc{AgenticSCR} operates, specifying what the agent can observe, read, execute, and modify. 
It comprises:

\begin{itemize}
\setlength{\leftskip}{-10pt} 
    \item \emph{CLI environment}: the execution host for running \textsc{AgenticSCR}.
    \item \emph{Filesystem scope}: a working directory corresponding to the Git repository under review.
    \item \emph{Toolchain}: a constrained set of executable command-line tools that enable \textsc{AgenticSCR} to interact with the environment (e.g., \texttt{git}, \texttt{grep}, \texttt{ls}).
\end{itemize}

In practice, repositories may be large, while the context window of Large Language Models (LLMs) is limited (e.g., up to 200{,}000 tokens for Claude).\footnote{\url{https://platform.claude.com/docs/en/build-with-claude/context-windows}}
To enable efficient perception, \textsc{AgenticSCR} adopts a \emph{diff-centric} strategy. 
It leverages version control introspection (e.g., \texttt{git diff}) to focus on modified files and relevant line ranges, rather than ingesting the entire repository. 
This prioritizes recent code changes, which are strongly associated with vulnerability introduction~\cite{lomio2022jitvd, fu2022linevul}. 
Unlike static LLM prompt-based approaches~\cite{tang-etal-2024-codeagent}, the perception component continuously gathers and integrates contextual information, enabling the agent to reason over both code changes and broader repository context.

\subsection{Chain of Subagents and its Policy and Goals ($\mathcal{A},\Pi, \mathcal{G}$)}


AgenticSCR is structured as a sequential composition of specialized subagents, each responsible for a distinct role. 
\rev{This is motivated by the complementary nature of detection and validation: conflating both objectives within a single agent risks optimizing for pattern recognition at the expense of precision. }
It adopts a detector--validator design pattern, consisting of a detector subagent ($a_d$) and a validator subagent ($a_v$). 
The \emph{detector} identifies potentially immature vulnerabilities, while the \emph{validator} assesses whether these findings constitute well-defined security issues. 
The coordination policy ($\Pi$) is defined through an explicit instruction---``\texttt{Use the `detector-subagent` to find security issues, then use the `validator-subagent` to validate the code review comments}''---which enforces a structured workflow. 
This enables AgenticSCR to manage the complexity of secure code review, requiring both syntactic pattern recognition and semantic validation.

\subsubsection{\textbf{Detector Subagent ($\mathcal{G}$, $a_d$)}}


The goal ($\mathcal{G}$) of the detector subagent ($a_d$) is to localize, detect, and explain vulnerabilities in early-stage code changes. 
As general-purpose LLMs lack systematic expertise in security analysis, we design the detector as a security-aware reviewer guided by a curated set of Static Application Security Testing (SAST) rules stored in long-term semantic memory.

SASTs provide a rich source of rule-based knowledge for detecting a wide range of coding issues.
Prior studies have shown that SAST can meaningfully assist code review.
For example, SAST warnings often overlap with developer comments, helping reduce review effort~\cite{Singh2017EvaluatingEffort}, and can prompt open-ended questions that lead to higher-quality code revisions~\cite{Taheri2020Similarity-basedFeatures}.
However, when applied purely statically, they often produce false positives due to the lack of contextual reasoning~\cite{Zhu2014SupportingAnalysis}. 
Instead of executing SAST rules directly, the detector leverages them as knowledge for reasoning, enabling more contextualized and precise analysis.

The detector operates as follows: (1) load SAST rules into memory; (2) identify relevant code changes via tool usage; (3) analyze changes with respect to the rules; and (4) reason about potential vulnerabilities. 
It produces three outputs:

\begin{itemize}
\setlength{\leftskip}{-10pt} 
    \item (i) \textbf{Line-level localization}, identifying relevant files and lines associated with potential vulnerabilities.
    \item (ii) \textbf{Vulnerability type}, expressed as CWE identifiers upon validation~\cite{mitreCWE}, aiding severity assessment and remediation~\cite{Christakis2016WhatStudy}.
    \item (iii) \textbf{Secure code review comments}, explaining the issue and supporting evidence from code context.

\end{itemize}

\begin{mdframed}[backgroundcolor=green!5, linecolor=green!60!black, linewidth=1pt, roundcorner=5pt, innertopmargin=5pt, innerbottommargin=5pt, skipabove=1em, skipbelow=10pt,  frametitle={\textbf{AgenticSCR's Detector Subagent}},
  frametitlealignment=\center,
  frametitlerule=true,frametitleaboveskip=0.6em,
    frametitlebelowskip=0.6em]
\small
\textbf{\# Detector Subagent}
You are an experienced secure code reviewer using Static Application Security Testing (SAST) methodology. Your task is to conduct a comprehensive security-focused code review by exploring the repo and generating actionable security comments following SAST rules and patterns.

\textbf{\#\# Workflow}

1. Load SAST Rules Resource

2. Find Diff Changes to Review

3. Examine Changes with SAST Focus

4. Apply SAST Rules to Changes

\textbf{\#\# Tools}

\textbf{\#\# Output Format}
\end{mdframed}
\vspace{1em}

\subsubsection{\textbf{Validator Subagent ($a_v$)}}


The goal ($\mathcal{G}$) of the validator subagent ($a_v$) is to assess the correctness and security relevance of review comments generated by the detector, filtering out false positives and retaining only substantiated vulnerabilities. 
While the detector may surface a broad range of potential issues, not all correspond to genuine security weaknesses, and unchecked noise can reduce developer trust and increase remediation effort~\cite{Johnson2013WhyBugs, Christakis2016WhatStudy}.

Coding weaknesses are a source of software vulnerabilities~\cite{Alfadel2023EmpiricalPackages, Charoenwet2024TowardWeaknesses}.
We hypothesize that supplying the validator subagent with concise, queryable descriptions of common coding weaknesses can improve its ability to recognize security-related issues in early-stage code changes.
We construct a lightweight knowledge base with a hierarchical organization inspired by the \textit{knowledge tree} concept~\cite{Li2024HierarchicalSurvey}.
The structure is derived from the Common Weakness Enumeration (CWE-1000) taxonomy~\cite{CWE1000}, from which we extract relevant weakness categories, short descriptions, and representative examples, providing agents with a lookup mechanism to relate observed code patterns to known classes of weaknesses.


The validator operates as follows: (1) load the CWE knowledge base into memory; (2) process candidate review comments; (3) map each comment to relevant CWE entries; and (4) apply CWE-based validation criteria (e.g., required conditions, exploitability, and context). 
Comments that do not satisfy these criteria are filtered out as likely false positives. 
The output is a refined set of validated review comments, each associated with a confirmed CWE weakness.

\begin{mdframed}[backgroundcolor=green!5, linecolor=green!60!black, linewidth=1pt, roundcorner=5pt, innertopmargin=5pt, innerbottommargin=5pt, skipabove=1em,  frametitle={\textbf{AgenticSCR's Validator Subagent}},
  frametitlealignment=\center,
  frametitlerule=true,frametitleaboveskip=0.6em,
    frametitlebelowskip=0.6em]
\small
\textbf{\# Validation Code Review}
You are an experienced secure code reviewer using the Common Weakness Enumeration (CWE) methodology. You will read a code review comment file (JSON format) from <target\_warning\_path> and validate each review comment using the CWE knowledge base. Filter out comments that do not correspond to real security issues or are likely false positives.

\textbf{\#\# Workflow}

1. Load CWE Tree Resource

2. Input Processing

3. CWE-Based Validation Process

4. CWE-Specific Validation Criteria

\textbf{\#\# Tools} 

\textbf{\#\# Output Format}
\end{mdframed}
\vspace{1em}

\subsection{Memory Model ($\mathcal{M}$)}


AgenticSCR is inspired by cognitive architectures that emphasize explicit memory and control in problem solving. 
Following prior work~\cite{sumers2023cognitive}, we model memory as $\mathcal{M} = \langle M_s, M_w, M_e \rangle$, consisting of semantic, working, and episodic memory.

\subsubsection{\textbf{Security-Focused Semantic Memory ($M_s$)}} 
\label{sec:semantic-memory}

Semantic memory is a type of long-term memory that stores reusable general knowledge, concepts, facts, rules, and meanings, independent of the specific context in which they were learned.
To compensate for the limited security specialization of general-purpose LLMs, we augment $M_s$ with structured security knowledge in both subagents.


\textbf{Static Application Security Testing (SAST) rules} are used by the detector subagent to identify vulnerability patterns. These rules encode well-established, empirically validated patterns derived from decades of security research and industrial practice (e.g., CWE, CERT, OWASP). 
By incorporating SAST rules, AgenticSCR grounds its analysis in authoritative security knowledge rather than relying solely on pretrained knowledge and probabilistic inference, thereby reducing hallucinations and improving alignment with established security definitions.

To incorporate SAST rules, we curate the CodeQL query suites\footnote{\url{https://github.com/github/codeql/}} into a structured and queryable document. 
We select CodeQL due to its permissive licensing and broad coverage of security-related issues.
Each rule is represented as a JSON object containing 
rule identifier,
a short \texttt{description}, and a high-level \texttt{pattern} summarizing what the rule detects.
When available, we record the mapped \texttt{cwe\_id} and \texttt{cwe\_name}, a \texttt{severity} label, \texttt{tags} (including CWE tags), structured \texttt{remediation} text, and paired \texttt{examples} showing vulnerable (\texttt{bad}) and secure (\texttt{good}) code.
For instance, the rule \texttt{py/timing-attack-against-hash}, mapped to CWE-208: Observable Timing Discrepancy, includes examples contrasting insecure equality checks with constant-time comparisons using \texttt{hmac.compare\_digest}.~\footnote{The complete rule in JSON format is provided in the replication package.}
By exposing these rule definitions in a machine-readable format, linking them to CWE identifiers, categories shared with the guideline, and taxonomy knowledge bases, we enable the detector subagent to (1) cross-check tentative findings against known rules, (2) reuse rule descriptions, and (3) generate natural-language review comments, while still exercising contextual judgment rather than blindly surfacing every static warning.

\textbf{CWE (Common Weakness Enumeration) tree}, used by the validator-subagent to reason about vulnerability types and relationships.
The CWE tree provides a canonical taxonomy that enables the validator subagent to reason about security findings at multiple levels of abstraction. 
Rather than treating detected issues as isolated code smells, the validator can situate each finding within a structured hierarchy of weaknesses (e.g., from high-level classes such as Improper Input Validation to concrete instances such as SQL Injection). 
This hierarchical reasoning allows the agent to check whether a detected pattern truly corresponds to a recognized security weakness, reducing misclassification and over-reporting.

To incorporate CWE taxonomy, we encode this taxonomy as a JSON \rev{covering 944 weakness entries derived from CWE-1000, organized into high-level categories} such as \texttt{injection}, \texttt{auth}, \texttt{crypto}, \texttt{input}, and \texttt{path}) and a list of weakness entries.
Each entry includes a CWE identifier (e.g., CWE-89), name, category, and subcategory (e.g., SQL injection), an estimated severity, and a shortened description that preserves the core weakness semantics.
To support retrieval from code context, we further attach patterns and \texttt{detection\_indicators} (e.g., \texttt{input\_validation}), relevant languages, and succinct examples with vulnerable and secure variants, plus remediation\_steps and a short hint such as "\texttt{Look for: string concatenation in SQL queries}".
As the CWE categories are aligned with the guideline categories, the agent can move from a general concern (e.g., input validation) to specific CWEs and reuse their standardized names, hints, and fixes when formulating review comments.

\subsubsection{\textbf{Built-In Working Memory ($M_w$) \& Episodic Memory ($M_e$)}} 


Both Working Memory ($M_w$) and Episodic Memory ($M_e$) are standard, built-in memory mechanisms within our agentic AI infrastructure. 
The primary purpose of working memory is to maintain short-term, transient, task-specific context during an active CLI session. 
On the other hand, episodic memory ($M_e$) captures reasoning traces and execution trajectories from past review runs (aka. session logs). 
These logs are persisted via agentic infrastructure to provide a high-fidelity record of how vulnerabilities were identified and validated.



\subsection{Tool Use ($\mathcal{T}$)}


AgenticSCR enables subagents to invoke external tools during analysis, including repository inspection utilities, diff analyzers, and command-line interfaces. 
\rev{Tool use is determined autonomously by the agent based on its current reasoning state rather than a fixed invocation order, grounding reasoning in actual program artifacts.}
Outputs from tool calls are stored in working memory ($M_w$) and reused across reasoning steps, supporting iterative and context-aware analysis. 
This interaction model allows the system to operate on realistic development contexts, such as incremental code changes, without exceeding the LLM context window.

\begin{itemize}
\setlength{\leftskip}{-10pt} 
    \item \texttt{open\_files}: load source files and security resource files.
    \item \texttt{expand\_code\_chunks}: selectively expand diff hunks and surrounding context.
    \item \texttt{grep}: locate security-relevant patterns in the repository.
    \item \texttt{expand\_folder}: inspect directory-level structure when needed.
    \item \texttt{bash} execute controlled shell commands (e.g., \texttt{git diff}, \texttt{git remote show origin}, \texttt{git diff <base>...HEAD}).
\end{itemize}

\subsection{Implementation}
In this paper, we use Atlassian's Rovo Dev CLI~\cite{RovoCLI2025} as a scaffold for implementing AgenticSCR.
\rev{While instantiated on Rovo Dev CLI, the framework is designed to be platform-agnostic and generalizable to other agentic infrastructures.}
We use \textbf{Claude Sonnet 4} as a base reasoning model for our agent.
Concretely, we implement the subagents (i.e., detector and validator), and the security-focused semantic memory (i.e., SAST rules and a CWE-based knowledge base).
Then, we integrate them into Rovo Dev CLI, resulting in AgenticSCR.  
During CLI invocation, AgenticSCR manages a transient \emph{working memory} ($M_w$) that contains the current diff, expanded code hunks, and intermediate hypotheses; this context is passed implicitly between tool calls and subagent steps and is discarded once the review completes.
In parallel, AgenticSCR records reasoning traces and tool trajectories as \emph{episodic memory} ($M_e$), which we can use for offline inspection, debugging, and analysis of subagent behavior.

\section{Experimental Setup}





\subsection{SCRBench Dataset Creation}


AgenticSCR is designed to operate in a local desktop environment for early-stage secure code review.
Accordingly, evaluating AgenticSCR requires: (1) public CVE vulnerabilities\footnote{Common Vulnerabilities and Exposures (CVE) is a public catalog of reported exploitable vulnerabilities with associated metadata, including CWE classifications.} with associated CWE types;\footnote{Common Weakness Enumeration (CWE) is a community-developed taxonomy of software and hardware security weaknesses.}
(2) sandbox environments simulating local Git repositories with vulnerability-introducing pre-commit changes; and (3) human-verified line-level vulnerability ground truths.
To satisfy these requirements, we construct a new benchmark dataset, \textsc{SCRBench}.
SCRBench contains: (1) a Git repository providing repository-level context; (2) staged early-stage code changes as input to AgenticSCR; (3) human-verified vulnerable lines for localization evaluation; and (4) associated CWE types and CVE descriptions for evaluating secure review comments and vulnerability classification.
Below, we describe the dataset construction process.

\subsubsection{Data Collection and Filtering Steps}

We begin with 3,256 public CVEs curated by Iannone et al.~\cite{Iannone2022TheStudy}.
Each CVE includes contains a CVE ID, an actual bug description, a bug-fixing commit (or CVE patches) of the associated repositories, and the associated CWE vulnerability types.\footnote{Example CVE: \url{https://nvd.nist.gov/vuln/detail/CVE-2017-16016}}
Following prior work~\cite{Costa2017SZZ}, we define the fixing commit as $c_n$ and the vulnerability-introducing commits as $\{c_{n-1},...,c_{n-m}\}$.
To ensure that each of the sandbox environments simulates a local working environment, we define the base commit ($c_{n-m-1}$) as a commit that happens \emph{before} the vulnerability is introduced in $c_{n-m}$ and is fixed in $c_n$.
To annotate which lines are actually vulnerable, prior studies~\cite{fu2022vulrepair, tang-etal-2024-codeagent} often regard the bug-fixing lines as the actual vulnerable lines for a given vulnerability. 
However, in real-world scenarios, the bug-fixing commit may contain other changes that are not related to the vulnerability, producing noise in the line-level ground-truths.
To mitigate such noise in the dataset, we manually verify which lines are truly relevant to a given vulnerability.
Given the large number of code changes, manual verification becomes infeasible; therefore, we select data based on the following criteria: 
(1) CVE vulnerabilities with at most three associated code changes; (2) code changes no larger than 112 modified lines (i.e., the median commit size before filtering); and (3) vulnerabilities affecting Python, JavaScript, or TypeScript, which are common programming languages at Atlassian.
Finally, we obtain 144 code commits spanning 107 CVEs, 92 GitHub repositories, 33 CWE types, and 3 programming languages.
Dataset statistics are summarized in Table~\ref{table-dataset-overview}.
Notably, 34.6\% of CVEs involve multiple related commits, indicating that vulnerabilities often emerge gradually rather than as single, fully manifested defects.
Because CWE types are sparse and long-tailed, we group the 33 CWE types into five higher-level categories following CWE-1000: \textit{Injection}, \textit{Authorization}, \textit{Information Exposure}, \textit{Resource Management}, and \textit{Control Flow}.
Additionally, the vulnerable-line ratio of 12.9\% suggests that vulnerable code is interspersed among non-vulnerable changes, increasing the difficulty of accurate localization during early-stage review.

\subsubsection{Sandbox Environment Creation}

To create a sandbox environment, we clone a Git repository, then we checkout the earliest commit that introduced vulnerable lines $c_{n-m}$ labeled by Iannone et al~\cite{Iannone2022TheStudy}.
To obtain the base commit ($c_{n-m-1}$) and code changes, we revert the commit $c_{n-m}$ into the \emph{uncommitted and unstaged} changes using the command \texttt{`git reset HEAD\textasciitilde1'}. 
Then, we execute \texttt{`git add .'} to convert all the unstaged changes (i.e., changed lines and files) into a Git staging area.
We ensure that all the changes in such a commit are properly staged by executing \texttt{`git diff'}.

\subsubsection{Human Verification of Line-Level Vulnerability Ground-Truths}


Not all lines modified in a bug-fixing commit correspond to the underlying vulnerability.
To enable accurate line-level localization and reduce noise, we manually verify which lines are associated with each CVE.
Specifically, we analyze the corresponding bug-fixing commit and identify vulnerable lines as those that (1) are added or modified from the base commit ($c_{n-m-1}$) and later modified or removed in the bug-fixing commit ($c_n$), and (2) align with the CVE description and associated CWE type.
This verification was conducted by the first and third authors, both experienced in security analysis during code review~\cite{Charoenwet2024TowardWeaknesses,Charoenwet2024AnReview}, \rev{with each decision cross-referenced against the CVE and CWE information to ensure consistency.}
Due to its complexity and reliance on manual judgment, the process required approximately seven days to complete.

\begin{table}[t]
\centering
\footnotesize
\caption{SCRBench dataset statistics, including the number of CVEs, CWE types, commits, and properties of code changes.}
\begin{tabularx}{\linewidth}{
    >{\hsize=1.5\hsize}X 
    >{\hsize=0.5\hsize}X
}
\toprule
\textbf{Property} & \textbf{Value} \\
\midrule
\# Vulnerabilities (CVEs) & 107 \\
\# Introducing Commits & 144 \\
\# Fine-Grained CWE Types & 33 \\
\# High-Level CWE Categories & 5 \\
\midrule
\% Vulnerabilities with Multi-Commit Introduction & 34.6\% \\
Code Changed Lines per Commit (median) & 54 \\
\bottomrule
\dag~The average proportion of vulnerable lines per commit.
\end{tabularx}
\label{table-dataset-overview}
\vspace{-5mm}
\end{table}




\subsection{Baselines}

We compare AgenticSCR against five baselines: a multi-agent code reviewer, a zero-shot static LLM reviewer, and three SAST tools.
Prior work has proposed various DL- and LLM-based approaches for secure code review~\cite{Liu2025SecureReviewer:Fine-tuning, Cihan2025EvaluatingReview, Haider2024PromptingGeneration, Peng2025ICodeReviewer:Prompts, Sun2025BitsAI-CR:Practice}. 
However, many of these approaches rely on fine-tuned models, proprietary training data, or system configurations that are not publicly available, making direct and reproducible comparisons infeasible.
\rev{Due to the lack of unified state-of-the-art early-stage secure code review, i.e., existing approaches target different settings, granularities, and datasets,}we select baselines that can be implemented using publicly available methods and artifacts:
\emph{Multi-agent code reviewer~\cite{tang-etal-2024-codeagent}:} We implement a role-based multi-agent reviewer following the framework of~\citet{tang-etal-2024-codeagent}.
\emph{Static LLM-based secure reviewer~\cite{Liu2025SecureReviewer:Fine-tuning}:} We implement a static LLM-based reviewer using the prompt provided by~\citet{Liu2025SecureReviewer:Fine-tuning}.
Both baselines use Claude Sonnet 4 as the backend model.
\emph{SAST tools:} We include three widely used tools: CodeQL (v2.23.3), Semgrep (v1.140), and Snyk (v1.130).

\subsection{Experimental Design}

We evaluate AgenticSCR and all baselines using 144 commits from SCRBench in a controlled sandbox environment. 
Each sandbox consists of a working directory containing staged but uncommitted code changes from the target commit.
Within each sandbox, we run AgenticSCR, both LLM-based baselines, and all three SAST tools under identical repository conditions.
AgenticSCR autonomously invokes command-line tools to identify relevant code changes (i.e., Step 2 in the Detector Subagent) and perform secure code review.
For the multi-agent and static LLM-based reviewers, we provide the full code changes via \texttt{git diff}. 
For SAST tools, we execute each tool at the repository level and filter out warnings unrelated to the target code changes at the file level, retaining only relevant results for evaluation.
To enable consistent comparison, we standardize outputs by extracting (1) the reported line, (2) the generated comment or warning message, and (3) the associated CWE type.

\subsection{Experimental Goals and Evaluation Measures}

We evaluate each approach at the comment level across three core secure code review tasks:
(1) line-level localization~\cite{fu2022linevul,AIBugHunter2023},
(2) secure comment generation~\cite{Liu2025SecureReviewer:Fine-tuning}, and
(3) vulnerability type prediction~\cite{fu2023vulexplainer,AIBugHunter2023}.
Accordingly, we define the following measures:

\textbf{\% of Correct Localization (L).} 
Percentage of generated comments whose reported vulnerable line falls within the same file and within $\pm5$ lines of the ground-truth vulnerable location (i.e., within 5 lines before or after)~\cite{tantithamthavorn2026rovodev}

\textbf{\% of Type Correctness (T).} 
Percentage of generated comments whose predicted vulnerability type matches the ground-truth vulnerability at the high-level CWE category.
We adopt high-level categories rather than individual CWE entries due to the hierarchical and interrelated nature of the CWE taxonomy, where a vulnerability may reasonably correspond to multiple CWE items.
Following prior work~\cite{Li2023ComparisonJava}, we apply a hierarchical grouping strategy to determine whether the predicted CWE and the ground-truth CWE belong to the same category.

\textbf{\% of Relevant Comments (R).} 
Percentage of generated comments that are semantically aligned with the ground-truth vulnerability.
Although early-stage secure code review is practical and beneficial, it lacks readily available human-written reference comments.
As a result, traditional human-alignment metrics used in automated code review (e.g., BLEU or SecureBLEU~\cite{Liu2025SecureReviewer:Fine-tuning, tantithamthavorn2026rovodev}) are not applicable.
Instead, we employ an LLM-as-a-Judge~\cite{Li2025FromLLM-as-a-judge} in a zero-shot setting. The judge evaluates whether a generated comment is relevant to the ground-truth vulnerability—based on its CWE ID, CWE name, CWE description, and CVE description—and returns a binary decision (True/False).

\textbf{\% of Overall Correctness (L\&T\&R).} 
Percentage of generated comments that simultaneously satisfy correct localization, correct vulnerability type, and semantic relevance.

\begin{mdframed}[backgroundcolor=blue!5, linecolor=blue!60!black, linewidth=0.5pt, roundcorner=3pt, innertopmargin=3pt, innerbottommargin=3pt, skipabove=.7em,  frametitle={\textbf{System Prompt for Comment Relevance Judge}},
  frametitlealignment=\center,
  frametitlerule=true,frametitleaboveskip=0.4em,
    frametitlebelowskip=0.4em]
\small
Is the code review comment below relevant to the given groundtruth CWE-ID, CWE Name, CWE Description, and CVE Description?

\noindent \textbf{[Code Review Comment]}\\
\noindent \textbf{[CWE-ID]}\\
\noindent \textbf{[CWE Name]}\\
\noindent \textbf{[CWE Description]}\\
\noindent \textbf{[CVE Description]}\\
\noindent Only respond with either `True`, when comment is relevant, or `False` when it is not, without any explanation.

\end{mdframed} 
\vspace{2mm}

We evaluate performance at the comment level to control bias from differences in the number of generated comments. 
Task-level metrics favor approaches that generate more comments, where success may reflect greater coverage rather than correctness. 
In contrast, comment-level evaluation assesses each comment independently, enabling fair comparison 
across approaches.

\section{Experimental Results}

We answer the following three research questions.





\subsection*{RQ1: How well does AgenticSCR perform secure code review tasks?}

\textbf{\underline{Results.} 
17.5\% of AgenticSCR-generated comments are simultaneously correctly localized, assigned the correct vulnerability type, and relevant to the target vulnerability (L\&T\&R), outperforming the multi-agent reviewer, static LLM reviewer, and SAST tools by 12.3, 10.6, and 13.8 percentage points, respectively.}
Table~\ref{table-generated-comments} summarizes the performance of all approaches across secure code review tasks.
AgenticSCR achieves the highest overall correctness (L\&T\&R) and performs best on most individual metrics (L, T, and R), indicating that security-focused memory augmentation improves the effectiveness of agentic secure code review.
This result demonstrates that augmenting agentic secure code review with structured security knowledge and LLM-based reasoning (i.e., Rules+Agentic+Reasoning) enables more effective early-stage secure code review than both a multi-agent reviewer without security-focused memory and traditional SAST tools that rely solely on rule-based analysis (i.e., Rules only).
\rev{The detector–validator architecture prioritizes precision over recall—a deliberate trade-off given the inherent difficulty of detecting partially manifested, context-dependent vulnerabilities at this early stage.}

In addition, \textbf{AgenticSCR also produces 71.3\% ($\frac{606-174}{606}$) – 85\% ($\frac{1085-174}{1085}$) relatively lower incorrect comments when compared to multi-agent reviewer, static LLM, and SAST tools.}
Lower incorrect comments are critically important in practice: at Atlassian, excessive irrelevant comments increase developer burden, disrupt development velocity, and lead to alert fatigue—causing genuine vulnerabilities to be overlooked or ignored.

\begin{figure}[t]
    \centering
    \includegraphics[width=\linewidth]{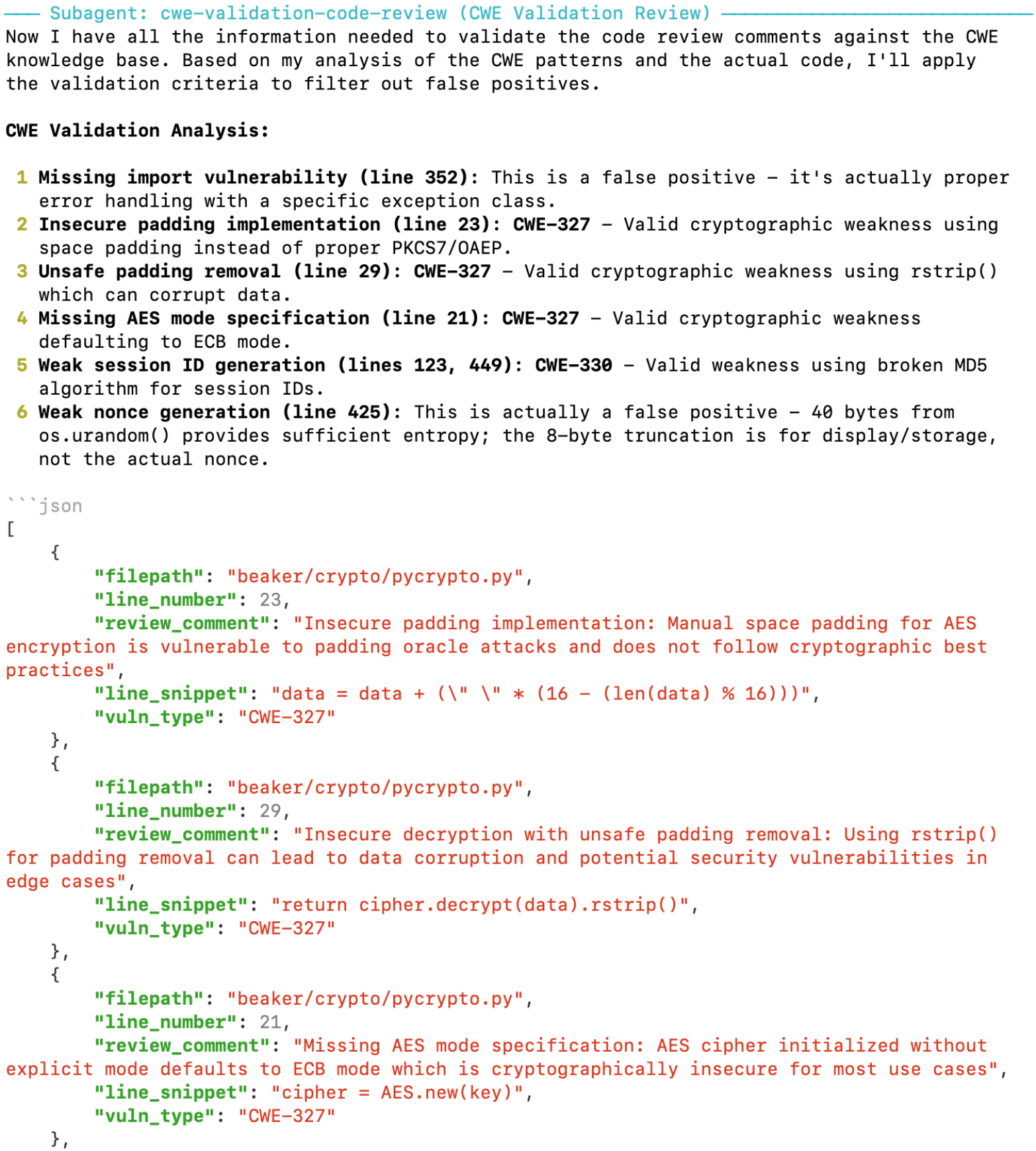}
    \caption{(RQ1) An example output of AgenticSCR, demonstrating the capability of the validator subagent filtering out irrelevant or invalid secure code review comments.}
    \label{fig:subagent-validation-screenshot}
    \vspace{-5mm}
\end{figure}

The reduction in incorrect comments appears attributable to the detector--validator architecture.
Figure~\ref{fig:subagent-validation-screenshot} illustrates an example in which the detector subagent generates six candidate review comments.
In this example, the detector subagent produces six candidate review comments. 
After validation, two comments (\#1 and \#6) are identified as false positives, leaving four comments (\#2–\#5) as final outputs.\footnote{Comment \#5 is omitted due to space limitations.}
Comparing the retained comments with the ground truth for CVE-2012-3458~\cite{CVE-2012-3458}
and its corresponding bug-fixing commit,\footnote{Bug-fixing commit: \url{https://github.com/bbangert/beaker/commit/c52d7b6ed1558d44acc96bd4f3468ffc4a8e5219}}
we observe that AgenticSCR correctly localizes and explains the vulnerability.
Specifically, AgenticSCR flags Lines 21 and 23, with Line 21 matching the annotated vulnerable location.
The generated comments identify improper encryption configuration that may expose observable patterns enabling inference of sensitive information.
The predicted vulnerability type, CWE-327 (Use of a Broken or Risky Cryptographic Algorithm), aligns with the ground-truth CWE-310 (Cryptographic Issues) under the same high-level category.

\begin{table}[t]
\caption{(RQ1) The performance of agentic code reviewers (AgenticSCR and Multi-Agent Reviewer), Static LLM, and SAST tools for secure code review tasks.}
\centering
\footnotesize

\begin{tabularx}{\linewidth}{
    >{\hsize=2.5\hsize}X |
    >{\hsize=0.7\hsize}Y |
    >{\hsize=0.7\hsize}Y
    >{\hsize=0.7\hsize}Y
    >{\hsize=0.7\hsize}Y |
    >{\hsize=0.7\hsize}Y
}

\toprule
\textbf{Approach} & 
\textbf{\#Cmt} &
\textbf{L} & 
\textbf{T} & 
\textbf{R} &
\textbf{L\&T\&R} 
\\
\midrule
\multicolumn{6}{c}{\textbf{Agentic AI for Secure Code Review}} \\
\midrule
\textbf{AgenticSCR} & \textbf{211} & \textbf{30.8\%} & 70.1\% & \textbf{63.5\%} & \textbf{17.5\%} \\
Multi-Agent Reviewer & 1,009 & 26.3\% & 16.4\% & 24.4\% & 5.2\% \\
\midrule
\multicolumn{6}{c}{Static LLM for Secure Code Review} \\
\midrule
Static LLM & 652  & 28.4\% & 44.0\% & 46.5\% & 6.9\% \\
\midrule
\multicolumn{6}{c}{Static Application Security Testing Tools} \\
\midrule
CodeQL 2.23.3 & 1,092 & 3.3\% & 19.2\% & 7.2\% & 0.6\% \\
Semgrep 1.140 & 376  & 5.1\% & \textbf{74.7\%} & 26.3\% & 3.7\% \\
Snyk 1.130 & 318 & 1.6\% & 39\% & 13.2\% & 0.9\% \\

\bottomrule
\end{tabularx}
\vspace{-5mm}
\label{table-generated-comments}
\end{table}

\subsection*{RQ2: Which vulnerability types does AgenticSCR perform best?}


Table~\ref{table:comment_relevance_by_group} presents performance across the five high-level vulnerability categories used in SCRBench.
For interpretability, we briefly summarize each category with representative examples below.

\begin{itemize}
\setlength{\leftskip}{-10pt} 

    \item \textbf{\underline{Inj}ection} (CWE-707): Missing input sanitization leading to injection vulnerabilities (e.g., CVE-2020-15141~\cite{CVE-2020-15141},
    arbitrary file write via path traversal).
    \item \textbf{\underline{Auth}orization} (CWE-287, CWE-284): Broken access control enabling privilege escalation (e.g., CVE-2015-4082~\cite{CVE-2015-4082},
    unauthorized modification of encryption settings).
    \item \textbf{\underline{Inf}ormation} (CWE-200): Exposure of sensitive data to unauthorized parties (e.g., CVE-2018-1000089~\cite{CVE-2018-1000089},
    forged email tracking via exposed logs).
    \item \textbf{\underline{Res}ource} (CWE-664): Improper resource management causing exhaustion or unsafe execution (e.g., CVE-2018-7889~\cite{CVE-2018-7889},
    code execution via malicious file loading).
    \item \textbf{\underline{Cont}rol} (CWE-691): Incorrect control flow or validation logic leading to unintended behavior (e.g., CVE-2020-11072~\cite{CVE-2020-11072},
    incorrect validation due to faulty comparison).
    
\end{itemize}

\noindent \textbf{\underline{Results.} 
Injection and Authorization are the two categories on which AgenticSCR performs best.}
AgenticSCR generates 23.4\% and 21.4\% correct comments (L\&T\&R) for Injection and Authorization vulnerabilities, respectively, outperforming all baselines in both categories.
These vulnerability types often involve recognizable security-sensitive patterns---such as missing input validation, unsafe sanitization, or absent access-control checks---that align well with AgenticSCR’s structured security knowledge and repository-aware reasoning.
For example, AgenticSCR correctly localizes and predicts the vulnerability type for code changes contributing to CVE-2015-1326~\cite{CVE-2015-1326}, an Injection-category vulnerability stemming from CWE-20 (Improper Input Validation).

\begin{table}[tbp]
\caption{(RQ2) The performance of AgenticSCR across the five high-level vulnerability types compared with other baselines.}
\centering
\footnotesize
\begin{tabularx}{\linewidth}{
    >{\hsize=2.2\hsize}X |
    >{\hsize=0.8\hsize}Y
    >{\hsize=0.8\hsize}Y
    >{\hsize=0.8\hsize}Y
    >{\hsize=0.7\hsize}Y
    >{\hsize=0.7\hsize}Y
}
\toprule
\textbf{Approach} &
\textbf{Inj.} &
\textbf{Auth.} &
\textbf{Inf.} &
\textbf{Res.} &
\textbf{Cont.} \\

\midrule
\multicolumn{6}{c}{\textbf{Agentic AI for Secure Code Review}} \\
\midrule

\textbf{AgenticSCR} &
\textbf{23.4\%} & \textbf{21.4\%} & \textbf{6.2\%} & \textbf{16.7\%} & 0.0\% \\

Multi-Agent Reviewer &
5.3\% & 0.1\% & 0.1\% & 2.1\% & 0.0\% \\

\midrule
\multicolumn{6}{c}{\textbf{Static LLM for Secure Code Review}} \\
\midrule

Static LLM &
12.6\% & 3.3\% & 1.5\% & 5.0\% & 0.0\% \\

\midrule
\multicolumn{6}{c}{\textbf{Static Application Security Testing Tools}} \\
\midrule
CodeQL 2.23.3 &
3.8\% & 0.0\% & -- & 0.0\% & 0.0\% \\
Semgrep 1.140 &
4.2\% & 0.0\% & 0.0\% & 2.5\% & 0.0\% \\
Snyk 1.130 &
2.6\% & 0.0\% & 0.0\% & 0.0\% & 0.0\% \\
\bottomrule

\end{tabularx}
\label{table:comment_relevance_by_group}
\vspace{-2mm}
\end{table}

In contrast, \textbf{AgenticSCR struggles on Information (6.2\%) and Control (0\%) vulnerabilities.}
For vulnerabilities in type Information, AgenticSCR may recognize potential issues in code changes, such as patterns where sensitive data flows into a risky sink, but incorrectly pinpoints the vulnerable lines.
For instance, in code changes contributing to CVE-2017-6200~\cite{CVE-2017-6200}, AgenticSCR predicts the correct vulnerability type and generates a relevant comment, but misidentifies the root vulnerable location.
\rev{Control vulnerabilities represent a fundamental limitation, with 0\% correctness across all approaches. 
These vulnerabilities arise from subtle logic flaws or violated business constraints—such as incorrect branching conditions or missing validation logic—that require domain-specific behavioral expectations and dynamic execution context unavailable from repository context alone.}

\subsection*{RQ3: What are the contributions of the security-focused components \rev{ and subagent design} in AgenticSCR?}

\begin{table}[t]
\caption{(RQ3) An analysis of the contributions of the security-focused components and subagent design of AgenticSCR.}
\centering
\footnotesize
\begin{tabularx}{\linewidth}{
    > {\hsize=3.9\hsize}X |
    >{\hsize=0.6\hsize}Y |
    >{\hsize=0.3\hsize}Y 
    >{\hsize=0.3\hsize}Y
    >{\hsize=0.45\hsize}Y |
    >{\hsize=0.45\hsize}Y
}
\toprule
\textbf{} &
\textbf{\#Cmt} &
\textbf{L} &
\textbf{T} &
\textbf{R}  &
\textbf{L\&T\&R}  \\

 \midrule
\multicolumn{6}{l}{\textbf{Base + SAST Rules and CWE Taxonomy (AgenticSCR)}} \\
\midrule
\textbf{Base + SAST Rules + CWE} &
\textbf{211} &
\textbf{30.8\%} &
\textbf{70.1\%} &
\textbf{63.5\%} &
\textbf{17.5\%} \\
Base + SAST Rules &
362 &
27.3\% &
57.7\% &
54.4\%  &
13.0\%  \\
Base &
1{,}070 &
23.4\% &
49.6\% &
31.0\% &
7.3\% \\
\midrule
\multicolumn{6}{l}{\textbf{+ Adding Secure Code Review Guidelines}} \\
\midrule
Base + Guideline &
757 &
24.2\% &
56.3\% &
49.3\%  &
10.3\%  \\
Base + Guideline + CWE &
371 &
26.7\% &
61.5\% &
58.2\%  &
12.4\%  \\
\midrule
\multicolumn{6}{l}{\textbf{+ Adding Both SAST Rules and Secure Code Review Guidelines}} \\
\midrule
Base + SAST Rules + Guideline &
604 &
28.1\% &
61.3\% &
51.2\%  &
12.6\%  \\
Base + SAST Rules + Guideline + CWE &
338 &
25.1\% &
63.3\% &
54.7\%  &
14.8\%  \\
\bottomrule
\end{tabularx}
\label{table:ablation-analysis}
\vspace{-5mm}
\end{table}

Table~\ref{table:ablation-analysis} presents the contribution of AgenticSCR's security-focused components, namely SAST rulesin the detector subagent and CWE Trees in the validator subagent.

\noindent\textbf{\underline{Results.}} Incorporating SAST rules yields an absolute improvement of 5.7\% over the base agentic CLI. 
Performance increases from 7.3\% (Base) to 13.0\% (Base+SAST), demonstrating that integrating SAST rules into long-term semantic memory strengthens the detector subagent's secure code review capability.


\textbf{Adding CWE Trees provides a further 4.5\% absolute improvement.}
Comparing Base+SAST+CWE with Base+SAST, performance increases from 13.0\% to 17.5\%, indicating that hierarchical vulnerability knowledge enhances the validator subagent’s ability to refine and validate candidate findings.
Despite generating fewer review comments overall (211 comments), AgenticSCR achieves the highest correctness, suggesting improved precision.
This reduction in noise likely stems from the CWE Tree enabling the validator to reason over vulnerability hierarchy, root causes, and contextual prerequisites rather than relying solely on flat pattern matching.

\rev{These results, combined with RQ1 where agentic repository exploration alone yields modest gains over static LLM prompting, disentangle the three contributing factors: agentic exploration provides contextual grounding, SAST rules drive detection quality, and CWE-based validation reduces noise.}

\section{Discussion}
In this section, we discuss the practical implications of AgenticSCR and further analyze key design choices underlying its performance and the impact of alternative security knowledge sources.

\underline{\textbf{Discussion 1:}} \emph{To what extent can AgenticSCR support secure code review at Atlassian?}
To further assess the practical effectiveness of AgenticSCR, we conduct a \textit{shadow deployment} experiment on an active project within Atlassian’s ecosystem.
Specifically, we select a Python- and JavaScript-based repository with 10--15 contributors, representing a typical engineering team.
We apply AgenticSCR to 64 intermediate commits associated with pull requests created between February and March 2026.
In total, AgenticSCR generates 41 security-related review comments across 28 commits, while the remaining commits receive no comments.
We then collaborate with two Product Security Engineers at Atlassian, each with at least three years of experience overseeing internal system security, to evaluate whether the generated comments warrant escalation to developers and provide justification for their assessment.
We note that this shadow deployment evaluates perceived practical usefulness rather than confirmed vulnerability discovery, as many reviewed changes were still under active development.





\begin{figure}[t]
    \centering
    \includegraphics[width=\linewidth]{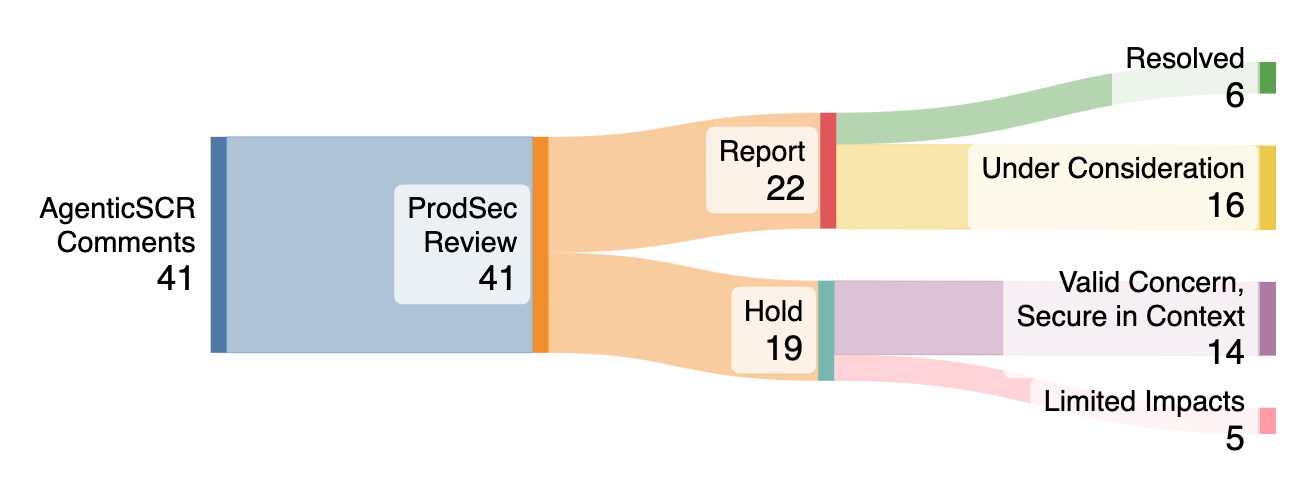}
    \caption{The Results of AgenticSCR's Shadow Deployment in Atlassian.}
    \label{fig:prodsec-review}
    \vspace{-6mm}
\end{figure}

As shown in Figure~\ref{fig:prodsec-review}, \textbf{54\% of generated comments (22 out of 41) raise security concerns that Product Security Engineers deem worthy of developer attention.}
Among these escalated comments, \textbf{27.3\% (6 out of 22) have already been \rev{addressed by developers}}, while the remainder are under consideration at the time of writing.
Of the non-escalated comments, Product Security Engineers assessed many as valid observations with limited practical impact or mitigated risk in the surrounding ecosystem.
For example, one comment flags potential sensitive-data exposure in logs; however, an engineer noted that an upstream dependency already sanitizes the data---an implementation detail unavailable to AgenticSCR.
These findings suggest that AgenticSCR can support real-world secure code review by surfacing \rev{candidate} security-relevant concerns and raising developers' awareness~\cite{Braz2021Why-} with a meaningful signal-to-noise ratio. 
Although some findings require ecosystem-specific knowledge, most generated comments were deemed at least technically valid by Product Security Engineers, indicating its practical value as an assistive review tool.

\underline{\textbf{Discussion 2:}} \emph{Do secure code review guidelines help our detector subagent?}
Various security standards and guidelines are commonly used in secure code review. 
Beyond SAST rules, such guidelines may serve as an alternative source of security knowledge for enhancing the detector subagent. 
To investigate this, we encode secure code review guidelines as persistent long-term semantic memory in AgenticSCR.
To avoid overwhelming the agent with verbose instructions—an issue noted in prior work~\cite{Pascarella2018InformationReview}—we distill code review guidelines from~\citet{Braz2022LessReview} and~\citet{Goncalves2025CodeTheories} into a compact, structured JSON format, where each entry encodes a category, essential checks, vulnerability patterns with examples, and remediation steps.
This representation guides the agent's attention to critical security patterns and fixes, enabling direct code reasoning while minimizing cognitive load and prompt overhead.

\textbf{The guideline contributes an absolute improvement of 3.2\%, but relatively lower than SAST rules by 24.3\% $(\frac{13.6 - 10.3}{13.6})$.}
Compared with the Base agent, Base+Guideline improves performance from 7.3\% to 10.5\%, demonstrating that code review guidelines also strengthen the detector subagent. 
Replacing SAST rules with guidelines (Base+Guideline+CWE) achieves only 12.4\% correct comments, 29.1\% $\left(\frac{17.5 - 12.4}{17.5}\right)$ lower than AgenticSCR. 
These findings confirm that while code review guidelines enhance reasoning, SAST rules provide stronger operationalizable signals.

\underline{\textbf{Discussion 3:}} \emph{Do both SAST rules and secure code review guidelines offer complementary capabilities to our detector subagent?}
It is possible that a combination of knowledge sources may improve the overall performance of AgenticSCR.
Thus, we conduct an additional analysis to investigate if the combination of both SAST rules and guidelines can improve the overall performance of AgenticSCR.

\textbf{The AgenticSCR variant augmented with both SAST rules and secure code review guidelines achieves 14.8\% overall correctness, which is 15.4\% $(\frac{17.5 - 14.8}{17.5})$ relatively lower than AgenticSCR.}
At the detector level, combining guidelines with SAST-rule memory leads to an absolute performance decrease of 0.4\%, from 13.0\% (Base+SAST) to 12.6\% (Base+SAST+Guideline without CWE validation).
This degradation may stem from the increased reasoning complexity and interference introduced by heterogeneous knowledge sources.
While SAST rules are pattern-driven and closely aligned with syntactic vulnerability manifestations, secure code review guidelines tend to be broader, more abstract, and more context-dependent.
Combining both within the detector subagent’s long-term semantic memory may therefore introduce semantic noise or competing signals, making it harder for the agent to prioritize actionable cues during detection.

\section{Threats to Validity}







\noindent\textbf{Internal Validity.}
Line-level vulnerability ground truths may be subject to annotation bias. 
To mitigate this, vulnerable lines must satisfy three criteria: (1) added or modified in pre-commit changes, (2) subsequently modified or removed in the bug-fixing commit, and (3) aligned with the CVE description and CWE type, with all decisions cross-referenced against CWE definitions.


\noindent\textbf{Construct Validity.}
Comment relevance relies on an LLM-as-a-judge~\cite{Li2025FromLLM-as-a-judge} to execute binary classification task, which may introduce model- or prompt-specific biases.
We validate judge reliability through manual evaluation on a \rev{human-labeled} set of 70 comments. 
Using GPT-4.1, the judge achieves an F1 score of 0.86 (Precision: 0.88, Recall: 0.86), indicating sufficient reliability for this task.

\noindent\textbf{External Validity.}
\rev{The filtering criteria reduce the original 3,256 CVEs (under 137 unique CWE IDs) to 107 CVEs across 33 CWE IDs.
All CWE IDs can be mapped onto the five high-level categories in SCRBench, suggesting representative coverage, though findings may not generalize to other languages or vulnerability types.}
The shadow deployment is limited in scale and duration and involves a small number of Product Security Engineers\rev{—a constraint imposed by organizational considerations}, which may limit the generalizability of the observed outcomes.
Therefore, we avoid strong generalization claims beyond the studied context.


\section{Conclusion}


\rev{We present \textsc{AgenticSCR}, an agentic secure code review support framework that assists human reviewers by surfacing candidate \emph{immature vulnerabilities} in early-stage code changes.}
\textsc{AgenticSCR} augments an agentic architecture with security-focused semantic memory, enabling subagents to autonomously access repository-level context while consulting SAST rules for detection and a CWE Tree for comment validation.
Our results show that \textsc{AgenticSCR} achieves the highest rate of correctly localized, relevant, and well-typed vulnerability comments (17.5\%), while generating 2--5$\times$ fewer comments than multi-agent reviewers, static LLMs, and SAST tools.
\rev{Moreover, 54\% of generated comments were deemed by Product Security Engineers as worthy of developer attention—not confirmed vulnerabilities—with 27.3\% of escalated comments already addressed by developers.}
These findings suggest that augmenting agentic AI with security-focused semantic memory can improve early-stage vulnerability detection, supporting shift-left security practices, while highlighting the remaining challenges in reliable localization, detection, and explanation of immature vulnerabilities.



\section*{Acknowledgment}
We would like to thank Kaif Ahsan and Jesse Merhi, Product Security Engineers at Atlassian, for their valuable insights and expert feedback in assessing the code review comments generated by AgenticSCR.
Kla Tantithamthavorn is the corresponding author.

\section*{Data Availability Statement}
The AgenticSCR framework, SCRBench dataset, and experimental and shadow deployment results are available at: \\
\url{https://doi.org/10.5281/zenodo.19908514}.

\bibliographystyle{ACM-Reference-Format} 
\bibliography{references,sample-base}

\end{document}